\documentclass[%
aps,
prapplied,
superscriptaddress,
longbibliography,
 amsmath,amssymb,
 reprint,%
]{revtex4-1}

\usepackage{graphicx}
\usepackage{dcolumn}
\usepackage{bm}
\usepackage{float}
\usepackage{array}
\usepackage{textcomp}
\usepackage{physics}
\usepackage[dvipsnames]{xcolor}
\usepackage{multirow}
\usepackage{booktabs}
\usepackage{upgreek}
\usepackage{xr}
\usepackage{cleveref}
\usepackage{tabularx}
\usepackage{amsmath}

\newcommand*{\citen}[1]{%
  \begingroup
    \romannumeral-`\x 
    \setcitestyle{numbers}%
    \cite{#1}%
  \endgroup   
}

\begin{document}
\newcolumntype{C}[1]{>{\centering\arraybackslash}p{#1}}
\newcommand{\beginsupplement}{%
        \setcounter{table}{0}
        \renewcommand{\thetable}{S\arabic{table}}%
        \setcounter{figure}{0}
        \renewcommand{\thefigure}{S\arabic{figure}}%
     }
     
\preprint{AIP/123-QED}

\title[Merged-element transmon]{Merged-element transmon}

\author{R.~Zhao}
\email{ruichen.zhao@nist.gov}
\affiliation{
Department of Physics, University of Colorado, Boulder, Colorado 80309, USA}
\affiliation{ 
National Institute of Standards and Technology, Boulder, Colorado 80305, USA
}%
\author{S.~Park}
\affiliation{
Department of Physics, University of Colorado, Boulder, Colorado 80309, USA}
\affiliation{ 
National Institute of Standards and Technology, Boulder, Colorado 80305, USA
}%

\author{T.~Zhao}
\affiliation{
Department of Physics, University of Colorado, Boulder, Colorado 80309, USA}
\affiliation{ 
National Institute of Standards and Technology, Boulder, Colorado 80305, USA
}%

\author{M.~Bal}
\affiliation{
Department of Physics, University of Colorado, Boulder, Colorado 80309, USA}
\affiliation{ 
National Institute of Standards and Technology, Boulder, Colorado 80305, USA
}%

\author{C.R.H.~McRae }
\affiliation{
Department of Physics, University of Colorado, Boulder, Colorado 80309, USA}
\affiliation{ 
National Institute of Standards and Technology, Boulder, Colorado 80305, USA
}%

\author{J.~Long}
\affiliation{
Department of Physics, University of Colorado, Boulder, Colorado 80309, USA}
\affiliation{ 
National Institute of Standards and Technology, Boulder, Colorado 80305, USA
}%

\author{D.P.~Pappas}%
\email{david.pappas@nist.gov}
\affiliation{ 
National Institute of Standards and Technology, Boulder, Colorado 80305, USA
}%
\date{\today}

\begin{abstract}

%

Transmon qubits are ubiquitous in the pursuit of quantum computing using superconducting circuits. However, they have some drawbacks that still need to be addressed. Most importantly, the scalability of transmons is limited by the large device footprint needed to reduce the participation of the lossy capacitive parts of the circuit. In this work, we investigate and evaluate losses in an alternative device geometry, namely, the merged-element transmon (mergemon). To this end, we replace the large external shunt capacitor of a traditional transmon with the intrinsic capacitance of a Josephson junction and achieve an approximately 100 times reduction in qubit dimensions. We report the implementation of the mergemon using a sputtered Nb--amorphous-Si--Nb trilayer film. In an experiment below 10 mK, the frequency of the readout resonator, capacitively coupled to the mergemon, exhibits a qubit-state dependent shift in the low power regime. The device also demonstrates the single- and multi-photon transitions that represent a weakly anharmonic system in the two-tone spectroscopy. The transition spectra are explained well with master-equation simulations. A participation ratio analysis identifies the dielectric loss of the amorphous-Si tunnel barrier and its interfaces as the dominant source for qubit relaxation. We expect the mergemon to achieve high coherence in relatively small device dimensions when implemented using a low-loss, epitaxially-grown, and lattice-matched trilayer.
\end{abstract}

\maketitle

\section{Introduction}
The invention of the transmon qubit has fueled the rapid development of quantum-information research over the past decade. Multiple landmark breakthroughs have been achieved with this technology. These include the mid-flight detection and reversal of a quantum jump~\cite{minev2019catch} and claims of the demonstration of quantum supremacy~\cite{arute2019quantum}. A variety of layout designs exist for transmons~\cite{minev2019catch,arute2019quantum,wu2017overlap,abdo2019active, weides2011coherence}, but all consist of a Josephson junction (JJ) with nonlinear inductance $L_{J}$, and a capacitor with capacitance $C$. In this architecture, one can exponentially suppress the charge dispersion of energy levels by increasing the $E_{J}/E_{C}$ ratio, where $E_{C}=e^2/2C$ is the charging energy and $E_{J}=\phi_0^2/L_{J}$ is the Josephson energy. Here, $\phi_0=\hbar/2e$ is the reduced flux quantum. On the other hand, the anharmonicity decreases linearly with increase in $E_{J}/E_{C}$~\cite{koch2007charge}. This unique property of transmons allows shielding of the qubit against the dephasing induced by the charge noise while maintaining a reasonably large anharmonicity. 

Nevertheless, transmon lifetimes can still suffer from qubit energy relaxation caused by parasitic two-level systems (TLSs) such as defects and dangling bonds that widely exist in superconducting circuits~\cite{muller2019towards,o2008microwave}. Several qubit architectures that are more resilient toward TLS losses have been developed in the past but require greater circuit complexity or an increased circuit footprint~\cite{manucharyan2009fluxonium,pop2014coherent,brooks2013protected,nguyen2019high}. Recent studies have demonstrated that it is possible to reduce the electric field participation ratio (PR) of the lossy materials in a transmon circuit by shunting the JJ with a large-scale coplanar capacitor~\cite{wang2015surface,gambetta2016investigating}. This mitigation strategy has significantly improved transmon coherence. However, the same design method also leads to a large, in-plane qubit footprint, fundamentally limiting the prospect of two-dimensional (2D) transmon integration.

\begin{figure}
\includegraphics[width=\columnwidth]{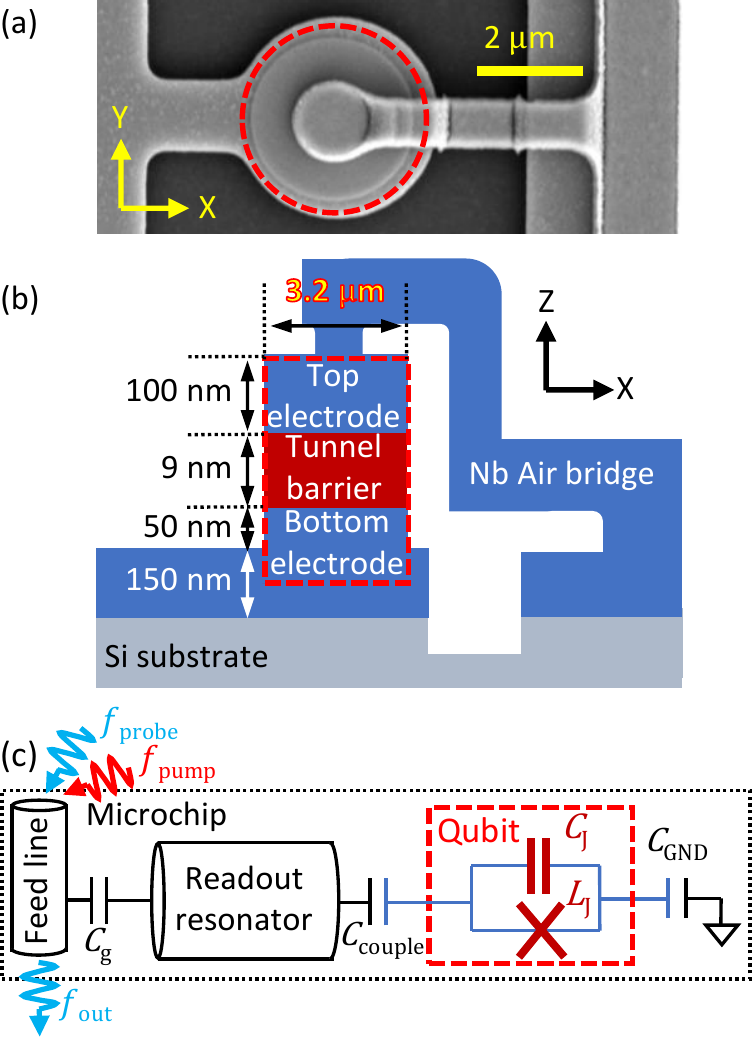}
\caption{(a) Scanning electron micrograph and (b) illustrative cross-section diagram of the mergemon. (c) The simplified circuit diagram of the microchip that holds the mergemon. Two interdigitated capacitors, $C_{\textmd{couple}}$ and $C_{\textmd{GND}}$, electrically connect the mergemon to the peripheral circuits and ground (GND). The output rf tone, $f_{\textmd{probe}}$,  of a VNA probes the response of the readout resonator. A second rf tone, $f_{\textmd{pump}}$ is used to populate the different energy levels of the mergemon circuit. All results presented in this paper are collected at a temperature below $10$~mK. The dashed red regions highlights the mergemon qubit. For ease of illustration, the dimension along the z axis is not to scale.\label{fig:fig1}}
\end{figure}

Here, we propose an alternative approach to scaling these circuits that minimizes qubit size while providing an avenue to significantly reduce losses due to interfaces, surfaces, and radiation. This design entails engineering the junction itself to satisfy the transmon requirements for frequency and anharmonicity by merging the external shunt capacitor and the JJ inductance into a single element made of a superconductor--tunnel-barrier--superconductor trilayer; that is, the ``mergemon.'' This design has several advantages over the traditional transmon. First, the mergemon allows a reduction of approximately 100 times in the device footprint~\cite{gambetta2016investigating,arute2019quantum}. Second, the resulting small qubit dimensions effectively suppress unwanted radiation and interqubit coupling through direct interactions or box modes. Third, the mergemon frequency could be less susceptible to the variation in lithography because the associated capacitive and inductive contributions towards the qubit frequency cancel out to first order.  Moreover, one may choose a low-barrier-height material as the junction tunnel barrier. This enables the use of a relatively thick tunnel barrier that may reduce the percentage variation in junction inductance. Finally, by leveraging the advanced molecular-beam-epitaxy process, tunnel barriers can be grown with atomic-level precision. Therefore, this design may achieve greater precision in qubit frequency allocations compared with the traditional transmons that use JJs fabricated via an \textit{in situ} oxidation process.

\section{device architecture and experimental methods}
For this proof-of-principle demonstration, we present the design, fabrication and measurement of the first-generation mergemon using a sputtered Nb--\textit{a}-Si--Nb trilayer. The microchip layout adopts a conventional coplanar-waveguide design~\cite{abdo2019active,solgun2019simple}. Figure~\ref{fig:fig1}(a) and (b) show a scanning electron micrograph and an illustrative picture of the cross section of the device. To achieve a balance between the charge-noise immunity and anharmonicity in the mergemon design, the $E_{\textmd{J}}/E_{\textmd{C}}$ ratio is set to 61. $E_{\textmd{J}}$ is calculated from the critical current predicted by the formula in Ref.~\citen{smith1982sputtered}. The $E_{\textmd{C}}$ is first approximated on the basis of a simple parallel-plate-capacitor model that considers only the junction capacitance $C_{\textmd{J}}$. Later, a circuit quantization process based on a simulated capacitance matrix fine-tunes $E_{\textmd{C}}$~\cite{devoret1995quantum}. The qubit frequency and anharmonicity are designed to be $5$~GHz and $260$~MHz, respectively. 

The fabrication process begins by sputtering a Nb/\textit{a}-Si/Nb trilayer onto a high-resistivity intrinsic Si wafer substrate that has been cleaned with hydrofluoric acid~\cite{olaya2009amorphous}. The mergemon structure and the peripheral circuitry are defined in two steps using optical lithography and a fluorine-based dry etch~\cite{olaya2009amorphous}. Because of the small (approximately $9$ nm) spacing between the top and bottom electrodes, the mergemon implements two interdigitated capacitors for coupling to the rest of the circuit, represented as $C_{\textmd{couple}}$ and $C_{\textmd{GND}}$ in Fig.~\ref{fig:fig1}~(c). We deposit and etch a Nb air bridge with a SiO$_x$ spacer that connects the top electrode to the interdigitated capacitors, and then dip the device in hydrofluoric acid to strip SiO$_x$ from the wafer.  

\begin{figure}[]
\includegraphics[width=0.8\columnwidth]{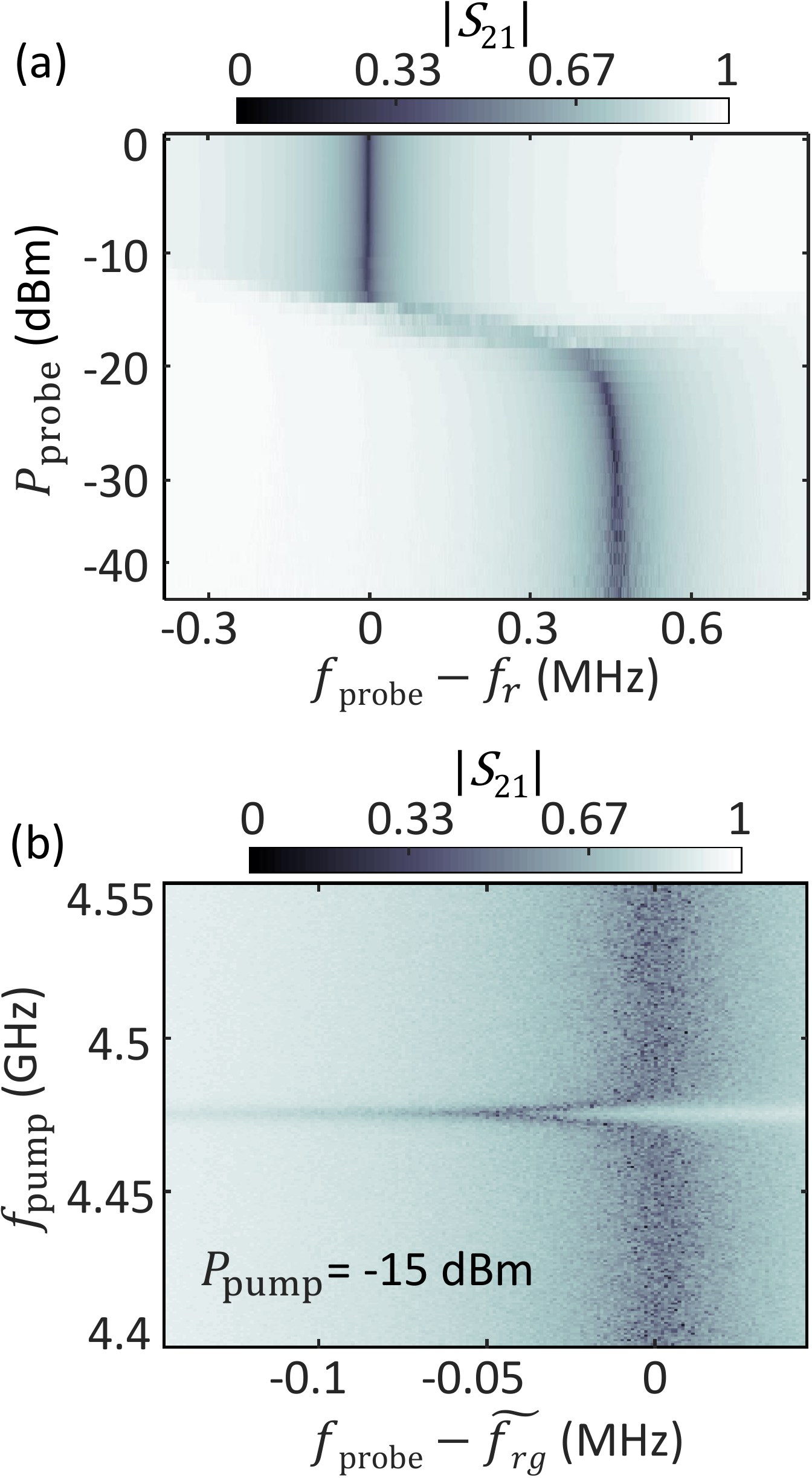}
\caption{(a) Transmission amplitude $|S_{21}|$ of the readout resonator plotted as a function of frequency and power of the probe tone. For clarity, the vertical axis has been offset by the bare readout resonator frequency $f_{r} = 6.876\,331$~GHz. (b) Transmission amplitude $|S_{21}|$ of the readout resonator plotted as a function of probe-tone frequency ($f_{\textmd{probe}}$) and pump-tone frequency ($f_{\textmd{pump}}$). The probe-tone power is set at $-40$~dBm to maintain dispersive coupling between the mergemon and the resonator. The change in the resonator frequency around $f_{\textmd{pump}}\approx4.475$~GHz represents the mergemon qubit-state transition. The horizontal axis is offset by the dressed resonator frequency for the qubit ground state $\widetilde{f_{rg}}=6.876\,796$~GHz.}
\label{fig:fig2}
\end{figure}

The microchip circuit described in Fig.~\ref{fig:fig1}(c) is characterized in a cryogen-free dilution refrigerator with base temperature below $10$~mK. Two microwave tones, $f_{\textmd{probe}}$ from port 1 of a vector network analyzer (VNA) and $f_{\textmd{pump}}$ from a microwave signal generator, are combined at room temperature and fed into the input port of the dilution refrigerator. The attenuators installed at various temperature stages filter the input signal and prevent the thermal noise from reaching the sample. The output signal $f_{\textmd{out}}$ is amplified at 4 K and room temperature before it reaches port 2 of the VNA. 

\section{results}
\subsection{Readout scheme and qubit frequency}
To facilitate the detection of the qubit state, a half-wavelength readout resonator is capacitively coupled the mergemon. The qubit-resonator coupling strength, $g/2\pi\approx50$~MHz, is much smaller than the detuning, $\Delta/2\pi = |f_{r}-f_{q}|$, where $f_{r}$ ($f_q$) is the resonator (mergemon) frequency. With this configuration, the qubit-resonator system is well described by the circuit quantum electrodynamics in the dispersive regime. The dispersive coupling between the qubit and the readout resonator induces a qubit-state-dependent shift of $\chi\sigma_z$ in the resonator frequency, where $\chi$ is the dispersive shift~\cite{wallraff2005approaching}. Figure~\ref{fig:fig2}(a) plots the transmission amplitude $|S_{21}|$ as a function of probe frequency and power. The probe power controls the number of photons in the resonator. For this measurement, the pump tone is turned off to keep the qubit in the ground state. The feature in the low-power regime ($P_{\textmd{probe}}<-25$~dBm) represents the dressed resonator frequency, $\widetilde{f_r}$. As the probe-tone power increases, the qubit-resonator system first enters a bifurcation regime where no clear resonance signals are observed. Finally, a resonance is observed approximately at the bare resonator frequency $f_r$ as the probe-tone power is further increased~\cite{bishop2010response}. This transition confirms the existence of coupling between the qubit and the resonator.

Next, the mergemon qubit frequency is examined in a two-tone spectroscopy. Similarly to the conventional transmons, the mergemon is a weakly anharmonic oscillator. The two-lowest energy states of the mergemon can be approximated as a qubit because of its finite anharmonicity. Therefore, to characterize the qubit system, one needs to choose a sufficiently small pump-tone power to induce the mergemon qubit transition while avoiding excitation to the higher energy states. Figure.~\ref{fig:fig2}(b) shows the probe-tone transmission amplitude $|S_{21}|$ for a pump tone of varying frequency. Visibly, the qubit transition causes a shift in the resonator frequency at $f_{\textmd{pump}}\approx4.475$~GHz.

\begin{figure}[b]
\includegraphics[width=0.8\columnwidth]{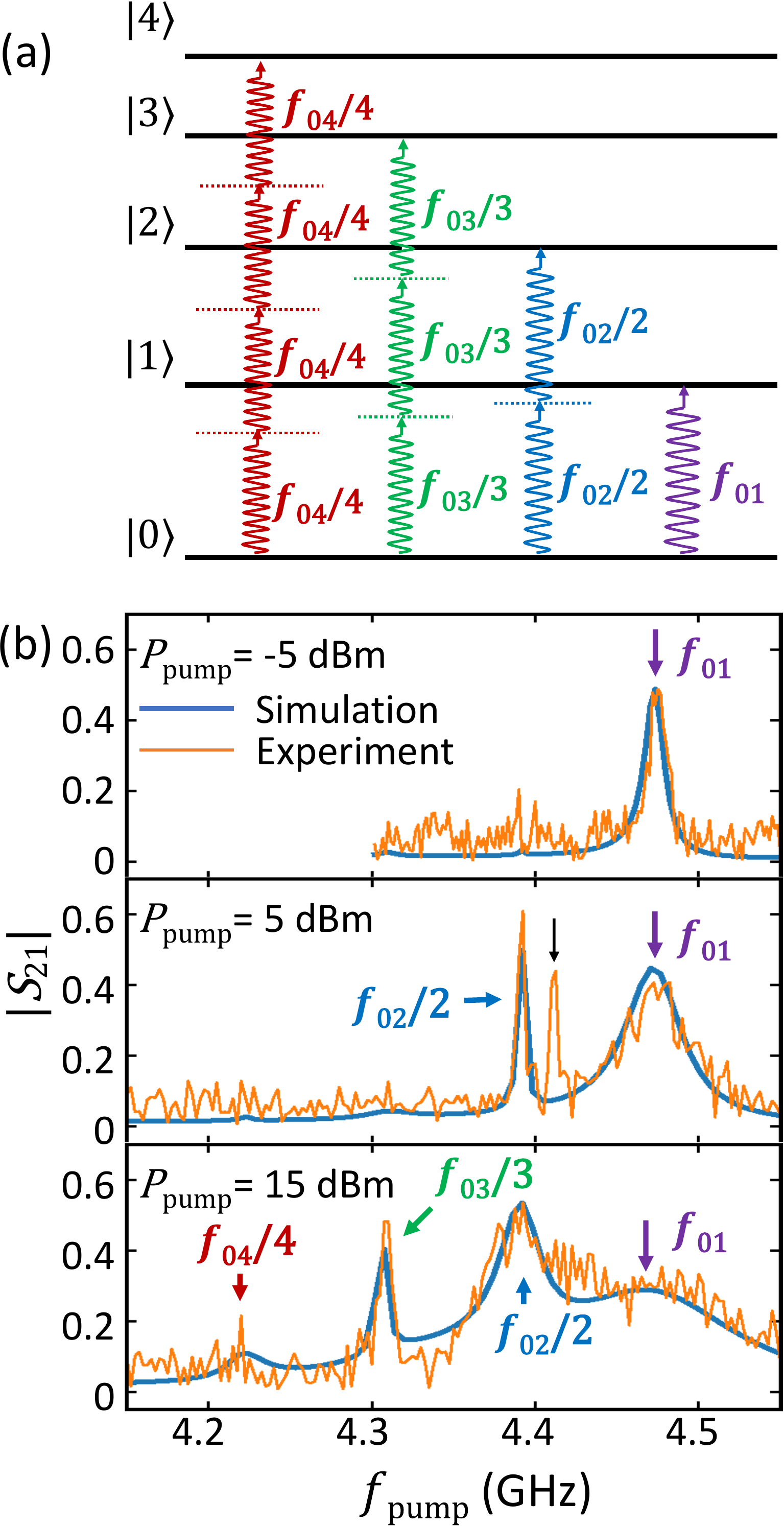}
\caption{(a) Illustrative diagram of the five-lowest energy states of the mergemon. The pump tone $f_{\textmd{pump}}$ drives the energy transitions between different states through single-photon or multiphoton processes. (b) Transmission amplitude $|S_{21}|$ of the readout resonator plotted as a function of pump-tone frequency $f_{\textmd{pump}}$. More energy states are populated as pump-tone power increases from $-5$ to $15$~dBm. Thick arrows highlight peaks corresponding to individual transition processes shown in (a). The thin black arrow points to a transition likely induced by a defect located in the \textit{a}-Si barrier. The probe tone is fixed at the frequency $\widetilde{f_{rg}}$ with sufficiently low power to avoid qubit excitation induced by a large resonator photon population~\cite{sank2016measurement}.}
\label{fig:fig3}
\end{figure}

\subsection{Mergemon energy-level structure and anharmonicity}
Now we turn to the investigation of the mergemon
energy-level structure and anharmonicity. Figure.~\ref{fig:fig3}(a) shows an illustrative diagram of the five-lowest energy levels of a mergemon. In this work, the anharmonicity $\alpha$ is defined as
\begin{equation}
\alpha=\frac{E_{\textmd{01}}-E_{\textmd{12}}}{\hbar},
\end{equation}
where $E_{\textmd{01}}$ ($E_{\textmd{12}}$) is the energy of the transition $\ket{0}\rightarrow\ket{1}$ ($\ket{1}\rightarrow\ket{2}$). To characterize the anharmonicity, the transitions to higher excited states are probed via the multiphoton processes described in Fig.~\ref{fig:fig3}(a). The spectra in Fig.~\ref{fig:fig3}(b) demonstrate that more states are populated as the pump-tone power increases. We extract the transition frequencies $f_{\textmd{01}}$ and $f_{\textmd{02}}$ from Fig.~\ref{fig:fig3}(b). Then, the anharmonicity can be calculated as
\begin{equation}
\alpha/2\pi=f_{\textmd{01}}-f_{\textmd{12}}=f_{\textmd{01}}-(f_{\textmd{02}}-f_{\textmd{01}})=2f_{\textmd{01}}-f_{\textmd{02}}.
\end{equation} 
The mergemon shows an anharmonicity of 170 MHz in the spectrum measurement, similar to that of a conventional transmon. Evidently, the measured mergemon frequency $f_q$ (4.475~GHz) and anharmonicity (170~MHz) significantly deviate from their designed values (5~GHz and 260~MHz). The scanning electron micrograph presented in Fig.~\ref{fig:fig1}(a) confirms there is no significant lithography error in the junction radius. Hence, the deviation in $f_q$ results primarily from the uncertainty in the calibration of the thickness of the \textit{a}-Si. The error in the barrier thickness is estimated to be $0.5$~nm. However, a lower measured qubit frequency implies that the tunnel barrier is thicker than the design. This, in principle, would give a smaller qubit capacitance and lead to a larger anharmonicity, which is in contradiction with the observed anharmonicity value. Hence, we conclude that the \textit{a}-Si relative permittivity used in our circuit design model contributes to the deviation in anharmonicity. For the \textit{a}-Si barrier,we assume a relative permittivity of 11.9 in the mergemon design, yet previous studies showed that the relative permittivity of \textit{a}-Si film can vary significantly from this value under certain growth conditions~\cite{brassard2003dielectric}. In our case, the measured qubit parameters project a relative permittivity of 17.50 for the sputtered \textit{a}-Si. Both errors can be corrected by more-precise calibration of the \textit{a}-Si growth rate and permittivity.

\subsection{Master-equation simulation}
To provide further evidence that the device is indeed in the transmon regime, the experimental result presented in Fig.~\ref{fig:fig3}(b) are compared with a master-equation (ME) simulation~\cite{johansson2012qutip,johansson2013qutip}. We follow the same simulation model and procedure that has
been used to study conventional-transmon circuitry~\cite{braumuller2015multiphoton}. The mergemon-resonator system can be described by the generalized Jaynes-Cummings Hamiltonian~\cite{jaynes1963comparison,koch2007charge,braumuller2015multiphoton}. In the basis of the isolated mergemon states $\ket{j}$, the Hamiltonian can be expressed as 
\begin{equation}
\hat{H} = \hbar\sum_{j}\omega_j\ket{j}\bra{j}+\hbar\omega_r\hat{a}^\dagger\hat{a}+\hbar\sum_{i,j}g_{ij}\ket{i}\bra{j}(\hat{a}^\dagger+\hat{a}),
\label{eq:JCH}
\end{equation}
where $\omega_j$ represents the eigenenergies of the isolated mergemon, $\omega_r=2\pi f_r$ is the angular bare resonator frequency, $\hat{a}^\dagger$($\hat{a}$) is the photon creation(annihilation) operator of the readout resonator, and $g_{ij}$ is the coupling matrix element between the transitions of the mergemon and the resonator. 

To find the dispersive shift in the resonator frequency induced by the mergemon, one needs to perform a canonical transformation on $\hat{H}$, omit second-order or higher-order terms, only consider coupling terms between nearest neighboring energy levels and apply a rotating-waveform approximation~\cite{braumuller2015multiphoton}. Then the angular dressed resonator frequency can be found as
\begin{equation}
\widetilde{\omega_{r}}=\omega_{r}-\chi_{01}\ket{0}\bra{0}+\sum_{j=1}(\chi_{j-1,j}-\chi_{j,j+1})\ket{j}\bra{j}).
\label{eq:dressedresonator}
\end{equation}
where $\chi_{j,j+1}=g_{j,j+1}^2/(\omega_{j,j+1}-\omega_r)$ is the dispersive shift induced by a transition between neighboring levels. The feedline $\abs{S_{21}}$ response, as shown in Fig.~\ref{fig:fig2}(b), demonstrates the absorption spectrum of the mergemon-resonator system and is suppressed at the probing frequency $f_{\textmd{probe}}=\widetilde{\omega_{r}}/2\pi$. Since the mergemon level excitation, $\langle n \rangle=\sum_jj\ket{j}\bra{j}$, induces a dispersive shift in $\widetilde{\omega_r}$, we may infer $\langle n \rangle$ from the $\abs{S_{21}}$ spectra in Fig.~\ref{fig:fig3}. The spectra are collected with a probing tone fixed at the dressed resonator frequency for qubit ground state $\widetilde{f_{rg}}$ so that the mergemon level excitation translates into the peak structure shown in Fig.~\ref{fig:fig3}.  

We note the isolated mergemon Hamiltonian takes the form
\begin{equation}
H_m = 4E_C(k-k_g)^2-E_J\cos{\phi},
\label{eq:charge1}
\end{equation}
where $k$ is the number of excess Cooper pairs on the island, $k_g$ is the offset charge and $\phi$ is the Josephson phase~\cite{koch2007charge}. One may rewrite the Josephson coupling term as $-E_J/2\sum_{k}\ket{k}\bra{k+1}+\ket{k+1}\bra{k}$ and diagonalize $H_m$. This would give the normalized coupling-matrix elements $g_{ij}/g_{01}$ and the eigenenergies of the isolated mergemon.
For these steps, $E_J$ and $E_C$ are derived from the mergemon design with minor corrections to compensate for the fabrication errors in the junction process. Using the resonator dispersive shift induced by the mergemon vacuum-state fluctuation $\chi_{01}$, as observed in Fig.~\ref{fig:fig2}(a), we calculate the coupling coefficient $g_{01}=\sqrt{\abs{\chi_{01}}(\omega_r-\omega_{q})}$.

By extending $\hat{H}$ to include the pumping tone $f_{\textmd{pump}}$, we derive the simulation Hamiltonian as
\begin{equation}
H_{\textmd{sim}} = \hat{H} + \hbar A\sum_{i,j} g_{ij}/g_{01} \ket{i}\bra{j}\cos{(2\pi f_{\textmd{pump}}t)},
\end{equation}
where $A$ is the pump-tone amplitude. The ME solver uses the Hamiltonian $H_{\textmd{sim}}$ to calculate the steady-state value for the level population, $\langle n \rangle$, of the anharmonic quantum circuit~\cite{johansson2012qutip,johansson2013qutip}. As shown in Fig.~\ref{fig:fig3}(b), the simulation is in good agreement with the experimental data for a wide range of pumping powers. 

\begin{table*}[]
\centering
	\caption{\label{tab1}Parameters for \textit{a}-Si mergemon TLS-loss analysis: material, relative permittivity ($\epsilon$), layer thickness ($d$), simulated participation ratio ($p_n$), TLS loss tangent ($\tan\delta$) from the literature, $\tan\delta$ reference and $T_1$ for each mergemon region. To take into account the interface defects between the electrode metal and the barrier, we assume $\tan\delta_{\textmd{TB-M}}$ is 10 times $\tan\delta_{\textmd{TB}}$. Similarly, considering the exposure of the tunnel-barrier--vacuum interface to harsh clean-room processes, $\tan\delta_{\textmd{TB-V}}$ is set to be 1000 times $\tan\delta_{\textmd{TB}}$. The vacuum region is lossless and, therefore, is omitted in this analysis.}

  \begin{tabularx}{\textwidth}{ccccccccc}
    \hline\hline
    Region            & Material  & $\epsilon$    & $d$ (nm) & $p_n$       & $\tan\delta$  & Reference                   & $T_1$~($\upmu$s)    \\ \hline
    Tunnel-barrier                   & \textit{a}-Si        & 11.9 & 9 & 7.01$\times10^{-1}$ & 5.00$\times10^{-4}$ & O’Connell \emph{et~al.}~\cite{o2008microwave}           & 9.08$\times10^{-2}$ \\ 
    Tunnel-barrier--Metal interface  & \textit{a}-Si/Nb     & 11.4 & 2    & 2.94$\times10^{-1}$ & 5.00$\times10^{-3}$ & $10\times\tan\delta_{TB}$   & 2.17$\times10^{-2}$ \\ 
    Tunnel-barrier--Vacuum interface & \textit{a}-Si/SiO$_x$& 4    & 2    & 8.03$\times10^{-5}$ & 5.00$\times10^{-1}$ & $1000\times\tan\delta_{TB}$ & 7.93$\times10^{-1}$ \\ 
    Metal-Vacuum interface          & Nb$_2$O$_5$ & 10   & 15   & 2.66$\times10^{-3}$ & 2.20$\times10^{-4}$ & Kaiser \emph{et~al.}~\cite{kaiser2010measurement}         & 5.45$\times10^{1}$  \\ 
    Substrate-Vacuum interface      & SiO$_x$     & 4    & 2    & 5.55$\times10^{-7}$ & 1.70$\times10^{-3}$ & Woods \emph{et al.}~\cite{woods2019determining}          & 3.37$\times10^{4}$  \\ 
    Metal-Substrate interface       & Nb/Si       & 11.4 & 2    & 5.73$\times10^{-7}$ & 4.80$\times10^{-4}$ & Woods \emph{et~al.}~\cite{woods2019determining}          & 1.16$\times10^{5}$  \\ 
    Substrate                       & Si          & 11.9 & $10^4$& 1.28$\times10^{-4}$ & 2.60$\times10^{-7}$ & Woods \emph{et~al.}~\cite{woods2019determining}         & 9.56$\times10^{5}$  \\
    \hline\hline
  \end{tabularx}

\end{table*}

\begin{figure*}[]
\includegraphics[width=1\textwidth]{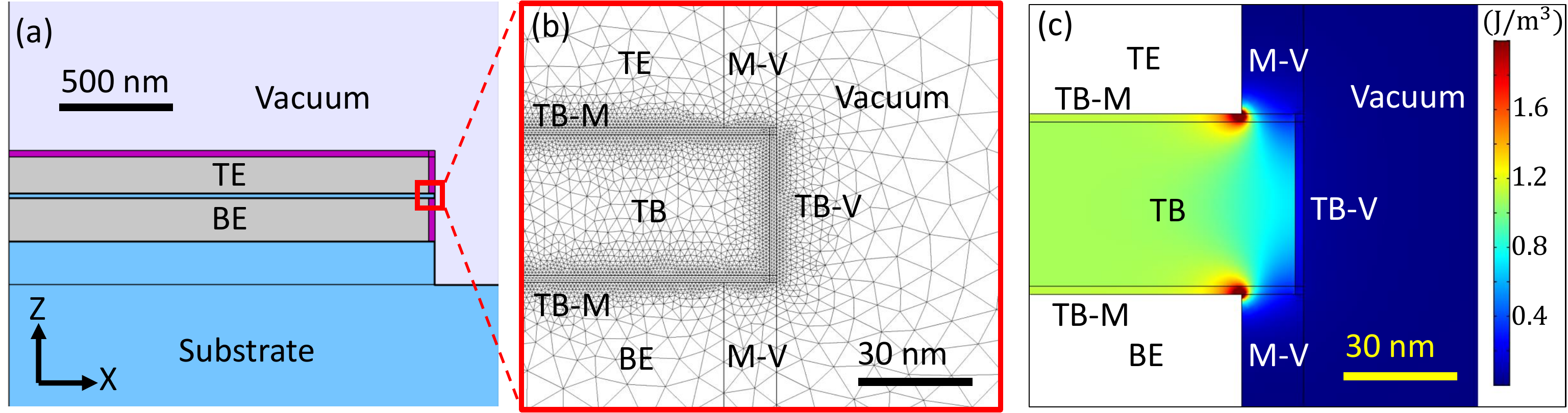}
\caption{(a)Cross-sectional images of a mergemon used in the finite-element simulation. The labels attached to each region illustrate the device partition used in the TLS-loss analysis. (b),(c) Enlargement of (b) the mesh grid and (c) the electric field energy profile near the tunnel-barrier (TB) interfaces. BE, bottom electrode; M, metal; TE, top electrode; V, vacuum.  }
\label{fig:fig4}
\end{figure*}

Since the spectrum linewidth is dominated by the qubit energy-relaxation process as discussed later, we set the dephasing rate in the ME simulation to zero. To achieve good agreement between the simulated and measured spectra, a qubit lifetime of $55$~ns is used in the ME simulation. We leave out the $f_{\textmd{probe}}$ in the simulation because it is detuned far from the mergemon frequency. Besides, we also carefully choose a small VNA output power to ensure its amplitude is negligibly small compared with $f_{\textmd{pump}}$. We estimate the probing tone reached at the mergemon chip is at least 2 orders of magnitude smaller than the pumping tone. Furthermore, our choice of low power for $f_{\textmd{probe}}$ and the input-line attenuation ensures the average number of photons in the readout cavity is around the single-photon level. Hence, the power-broadening effect due to the cavity photon shot noise should be negligible compared with the intrinsic line width of the mergemon.

\subsection{Two-level-system loss analysis}

Finally, we discuss the possible cause of the short qubit lifetime and strategies to increase it. The operation of a transmon qubit requires a small driving power close to a single-photon excitation. Hence, most of the parasitic TLSs in the superconducting circuit are unsaturated. These provide channels for energy relaxation, thereby limiting $T_{\textmd{1}}$ of the qubit~\cite{calusine2018analysis,wang2015surface,mcrae2020dielectric,o2008microwave}. In this work, we explore the PR model to understand TLS loss in the mergemon~\cite{wang2015surface}. For this analysis, the device is partitioned into regions listed in Table~\ref{tab1}. The PR is defined as
\begin{equation}
p_{n}=\frac{U_{n}}{U_{\textmd{total}}}=\frac{\int_{v_n}\frac{1}{2}\epsilon_{n}|E|^2}{\sum\int_{v_n}\frac{1}{2}\epsilon_{n}|E|^2} \label{eq:eq4}
\end{equation}
where $v_{n}$ is the volume of the $n$th region, $\epsilon_{n}$ is the relative permittivity of the $n$th region, $E$ is the electric field strength and $U$ is the electric field energy. Because of the difference in material and surface treatment, each region possesses a distinct TLS loss characterized by the loss tangent $\tan\delta_{n}$~\cite{mcrae2020dielectric}. Then, the qubit lifetime can be estimated as
\begin{equation}
T_{\textmd{1}} = \frac{1}{2\pi f_q\sum p_{n}\textmd{tan}\delta_{n}}. \label{eq:eq5}
\end{equation} 

We calculate the PR of each region using the electric field profile simulated in a finite-element field solver. First, we build the 2D model, shown in Fig.~\ref{fig:fig4}(a), that reflects the cross section of the mergemon. The thickness and relative permittivity of each region are listed in Table~\ref{tab1}. Figure.~\ref{fig:fig4}~(b) shows the mesh grid created for the 2D structure. A finer grid size is used for the thin interface regions. We use the 2D axisymmetric mode of the finite-element field solver, where we take advantage of the rotational symmetry of the model to obtain the three-dimensional field solution.  This approach allows more-efficient use of our computational resource. Then the integration of the electric field energy density, as shown in Fig.~\ref{fig:fig4}(c), over the volume of the respective regions yields $U_n$ used to calculate the PR presented in Table.~\ref{tab1}. Because of the high aspect ratio (more than $3000$) between the junction radius and the thickness of various interfaces, it is challenging to include the full-scale peripheral circuitry in the electric field simulation. Hence, the PR-model analysis focuses on the trilayer-junction region. Literature shows the coplanar resonators~\cite{woods2019determining,calusine2018analysis,mcrae2020materials,megrant2012planar}, capacitors~\cite{gambetta2016investigating} and air bridges~\cite{chen2014fabrication} contribute TLS losses much smaller than that of the a-Si barrier. Therefore, we believe this PR model should capture the major loss of the mergemon circuit. Using $\tan\delta_{n}$ of each of these regions, as reported in previous studies, we predict the mergemon $T_{\textmd{1}}$ to be 17.1~ns~\cite{calusine2018analysis}. This value deviates from the $T_{\textmd{1}}$ value estimated by the ME simulation mainly due to the measurement uncertainty of $\tan\delta_{n}$ reported in literature~\cite{mcrae2020materials,mcrae2020dielectric}. Nonetheless, this analysis estimates the order of magnitude of $T_{\textmd{1}}$ and, more importantly, provides valuable insights into the strategy to increase the mergemon lifetime.

\begin{figure}[]
\includegraphics[width=0.8\columnwidth]{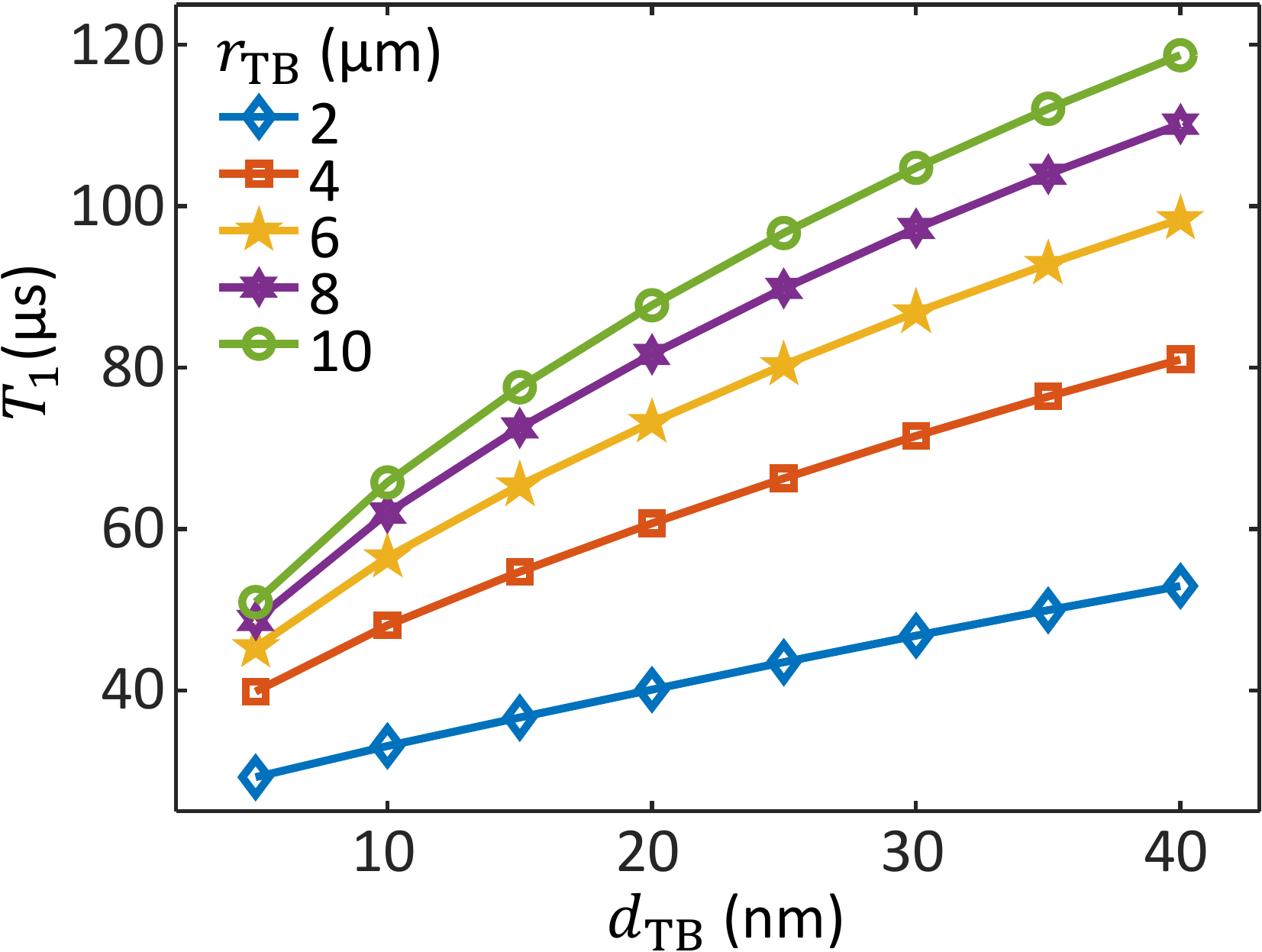}
\caption{Simulated $T_{\textmd{1}}$ of a mergemon with a crystalline tunnel barrier and interfaces plotted as a function of barrier thickness $d_{\textmd{TB}}$ and junction radius $r_\textmd{TB}$. For this analysis, we use $\tan\delta_{\textmd{TB}}$ of $1\times10^{-7}$, which is comparable to $\tan\delta$ of bulk crystalline silicon~\cite{woods2019determining}. The same scaling factor in Table~\ref{tab1} is used to derive $\tan\delta$ of tunnel-barrier--metal ($10\times\tan\delta_{\textmd{TB}}$) and tunnel-barrier--vacuum ($1000\times\tan\delta_{\textmd{TB}}$) interfaces. The loss tangents of all other interfaces and bulk materials are identical to the values presented in Table~\ref{tab1}.
}
\label{fig:fig5}
\end{figure}

To evaluate the contribution of individual materials and interfaces to the mergemon energy relaxation, $T_{\textmd{1}}$ of each region is derived with Eq.~\ref{eq:eq5}. As shown in Table~\ref{tab1}, the \textit{a}-Si mergemon lifetime is limited by the dielectric loss of the tunnel barrier and its interfaces. The PR model predicts the qubit lifetime can be increased to $33.2$~$\upmu$s by the replacement of the amorphous barrier and the associated interfaces (tunnel-barrier--vacuum and tunnel-barrier--metal) with high-quality crystalline material. Furthermore, our simulation reveals that further increase in the crystalline mergemon lifetime is possible through the modification of the device geometry along with the necessary material developments to realize these changes. By increasing the tunnel barrier thickness and radius, one can reduce the participation ratio of the lossy metal-vacuum interfaces at the edge of the tunnel barrier. Figure.~\ref{fig:fig5} shows the lifetime of a mergemon with a crystalline tunnel barrier can be prolonged beyond $100$~$\upmu$s with this approach. The footprint of the optimized mergemon is still approximately 20 times smaller than that of the conventional transmon~\cite{gambetta2006qubit,arute2019quantum}. 

Our results highlight the need for further studies in epitaxial trilayer material that would enable such an optimized mergemon design. The thick tunnel barrier suggests that low-band-gap semiconducting materials such as crystalline germanium could be a promising candidate for this application~\cite{smith1982sputtered}. In addition, the choice of electrode material and its lattice-matching condition with the tunnel-barrier could also impact the effective tunnel barrier height. We envision the low-loss trilayer material needs radical innovation in material manufacturing and process technology. Some initial progress on the trilayer growth has been reported recently~\cite{mcfadden2020epitaxial}. Because of the stringent loss requirement imposed by the mergemon architecture, it is also equally important to develop a high-accuracy TLS-loss-characterization technique that can reliably measure $\tan\delta$ on the order of $10^{-7}$ and has the capability to pinpoint the interface or material that causes the loss. One promising approach would be the dielectric loss extraction method recently reported in Ref.~\cite{mcrae2020dielectric}. We believe these two research efforts will be crucial for the discovery of the material that is suitable for implementation of a high-coherence mergemon qubit. 

\section{Summary and outlook}

In conclusion, we present a strategy to improve transmon scalability. By merging the shunt capacitor and the JJ into a single device made of a sputtered Nb/\textit{a}-Si/Nb trilayer, we achieve a 2-orders-of-magnitude reduction in the transmon footprint. The energy-level transitions measured by a two-tone spectroscopy confirm the mergemon is indeed a weakly anharmonic system. The two-tone spectra obtained with various pumping powers are described well with a master-equation simulation. An in-depth TLS-loss analysis identifies the lossy amorphous silicon tunnel barrier and its interfaces as the major limiting factor for the qubit relaxation time. We acknowledge that the large size of the readout resonator and the on-chip signal routing are still roadblocks for high-density qubit integration. However, these can be addressed by use of a lumped-element component~\cite{reagor2018demonstration} or by extension of the quantum-chip architecture into the vertical dimension~\cite{bejanin2016three}. Our analysis indicates the optimal host material for the mergemon qubit is a trilayer film grown by molecular-beam epitaxy. The lattice-matched molecular-beam-epitaxy-grown trilayer can significantly reduce the dielectric loss of the tunnel barrier and its interfaces. It also allows atomic-scale control over the barrier thickness that could enable high-precision qubit-frequency allocation. Both capabilities are crucial in building a large-scale quantum processor using mergemon qubits. 
\newline

\begin{acknowledgments}
We acknowledge the support of the NIST Quantum Initiative, the U.S. National Science Foundation (Grant No. 1839136), the Laboratory of Physical Sciences NEQST Program (Grant No. W911NF1810114) and the U.S. Department of Energy (Grant No. de-sc0019199). We thank D.~Olaya for enlightening discussions, A.E.~Fox and M.~Thompson for their assistance with device fabrication and D. Hite for valuable feedback on the manuscript. 
\end{acknowledgments}

\bibliography{Mergemon.bib}

\end{document}